# covSTATIS: a multi-table technique for network neuroscience

Giulia Baracchini[1*], Ju-Chi Yu[2*], Jenny Rieck[3], Derek Beaton[4], Vincent Guillemot[5], Cheryl Grady[3,6], Hervé Abdi[7], R. Nathan Spreng[1]

## Affiliations

[1] Montreal Neurological Institute, Department of Neurology and Neurosurgery, McGill University, Montreal, Canada
[2] Campbell Family Mental Health Institute, Centre for Addiction and Mental Health, Toronto, Canada
[3] Rotman Research Institute at Baycrest, Toronto, Canada
[4] Data Science & Advanced Analytics, Unity Health Toronto, Toronto, Canada
[5] Institut Pasteur, Université Paris Cité, Bioinformatics and Biostatistics Hub, Paris, France
[6] Departments of Psychiatry and Psychology, University of Toronto, Toronto, Canada
[7] School of Behavioral and Brain Sciences, The University of Texas at Dallas, Richardson, USA

*These authors contributed equally to this work

## Correspondence

Giulia Baracchini: giulia.baracchini@mail.mcgill.ca

# Abstract

Similarity analyses between multiple correlation or covariance tables constitute the cornerstone of network neuroscience. Here, we introduce covSTATIS, a versatile, linear, unsupervised multi-table method designed to identify structured patterns in multi-table data, and allow for the simultaneous extraction and interpretation of both individual and group-level features. With covSTATIS, multiple similarity tables can now be easily integrated, without requiring *a priori* data simplification, complex black-box implementations, user-dependent specifications, or supervised frameworks. Applications of covSTATIS, a tutorial with Open Data and source code are provided. CovSTATIS offers a promising avenue for advancing the theoretical and analytic landscape of network neuroscience.

# Main

Correlation, covariance and distance matrices are among the most commonly used data types in network neuroscience[1–4]. They are typically built via pairwise comparisons of functional and/or structural neuroimaging data. In these matrices, one entry stores a numerical value quantifying the similarity between two spatial locations (i.e., brain voxels, vertices, regions, channels) and the pattern of these entries reflects an estimate of brain network organization.

In network neuroscience, matrices—also called here data tables—are typically obtained from sets of variables collected on the same individuals (e.g., multiple scans or sessions, multiple imaging modalities)[5–9], or from the same variables collected on different individuals (e.g., one type of imaging scan on several participants)[10–13]. Data tables are then compared to one another to assess temporal network structure[14–17], multi-modal network organization[4,18,19], individual differences[20–25], and group or population effects[26–30]. Data tables are also contrasted with one another to investigate the statistical reliability of patterns derived from network neuroscience methods[31,32]. Similarity analyses among multiple data tables thus constitute the cornerstone of network neuroscience research.

In network neuroscience, similarity analyses are most often conducted on aggregate information within data tables (e.g., graph theory analyses[33]) or across (e.g., categorical groupings), or on single tables with reduced dimensions (e.g.,[34]). While these approaches reduce the high dimensionality of network neuroscience data and simplify the analytic landscape, they may obscure important properties of brain function that could be revealed from the analysis of full data tables. Approaches able to align and compare relational information from full data tables across multiple observations, can augment the utility, precision, and applicability of network neuroscience data and advance our understanding of brain network organization in health and disease.

Statistical methods exist— called multi-table methods[35–43]—explicitly designed for similarity analyses of full data tables. Their goal is to identify structured patterns within preserved high-dimensional multi-table data and explain and visualize their statistical dependencies. Multi-table methods serve as the basis of network investigations across scientific disciplines (e.g.,[44]), yet they are not well known in network neuroscience and therefore remain underused.

Network neuroscience presently counts a few multi-table methods, including machine learning and deep learning tools[45,46], graph neural networks[47,48], multi-layer and multiplex networks approaches[18,49], similarity network fusion techniques[50] and non-linear matrix decomposition algorithms[51–53]. These approaches have been applied to a variety of research questions about brain network organization both in health and disease[54–61]. Yet, they often yield complex results challenging to interpret, potentially because these methods rely on complex mathematical implementations[62,63] and supervised analytical frameworks[51] that do not allow results to be traced back to the original data. There is therefore a pressing need in network neuroscience for multi-table methods that preserve data fidelity and enhance interpretability. CovSTATIS solves this problem by analyzing intact data tables in a linear, unsupervised manner, thus allowing for the simultaneous extraction and interpretation of both individual and group-level features.

The covSTATIS method (and its variant DISTATIS) first appeared in[64] and is a three-way extension of multidimensional scaling and Principal Component Analysis[64–66]. The name, covSTATIS, combines "covariance" with "STATIS" (a French acronym for "structuring three-way statistical tables"). CovSTATIS takes as input symmetric, positive semi-definite matrices (i.e. symmetric matrices such as cross-product, covariance and correlation matrices) and assesses their similarity[67–69]. While covSTATIS is specifically designed for correlation/covariance matrices, there exists an equivalent approach for distance matrices called DISTATIS[64]. CovSTATIS and DISTATIS belong to the STATIS family of multi-table approaches. In neuroimaging, STATIS-based techniques have been applied in a limited capacity[70–76]. In network neuroscience, recent work from our group applied covSTATIS to compare spatial patterns of fMRI connectivity across task states[77], and to estimate resting-state fMRI connectivity dynamics[78].

In covSTATIS, the similarity among multiple correlation/covariance matrices is quantified by the $R_V$ coefficient—a matrix analogous to a squared correlation coefficient—which takes values in the interval $[0,1]$[67,79]. These coefficients are stored in an $R_V$ *similarity matrix* where each row/column corresponds to a single data table (**Figure 1, step 1**). Next, covSTATIS performs an eigenvalue decomposition (EVD) on the $R_V$ similarity matrix and takes the resulting first eigenvector to derive weights for each data table. Note that because the $R_V$ is always positive, all the entries for the first eigenvector are positive—a property derived from the Perron-Frobenius theorem. The first eigenvector maximally explains the variance in the $R_V$ matrix and quantifies how similar each table is to the common pattern. Weights for each data table are derived by scaling the first eigenvector to sum to 1. Higher weights identify tables that are more similar to the common pattern, whereas lower weights identify tables less similar to the common pattern. These weights are then used to linearly combine the data tables by multiplying each table by its weight and summing across all the weighted tables. This step generates a weighted group matrix—called the *compromise matrix* (**Figure 1, step 2**)—which is next decomposed by EVD. Orthogonal components are extracted from this second EVD and serve as the main output of covSTATIS (**Figure 1, step 3**).

Components reveal the similarity between variables with regards to the compromise—the ensemble of the similarity patterns across all data tables. For each component, *factor scores* are used to reconstruct and interpret covSTATIS' output in relation to the original data, yielding 1) *global factor scores* that represent associations between variables with respect to the compromise, and 2) *partial factor scores* that represent associations within each data table separately (**Figure 1, step 4**). For example, given a covSTATIS analysis of multiple functional connectivity matrices where only positive connectivity values are considered, *global factor scores* represent the brain regions on the component space and illustrate the associations in their connectivity profiles across the whole sample. *Partial factor scores* represent the brain regions of each individual's connectivity matrix and illustrate how regions are associated with each other in relation to the group pattern. Factor scores of any two components can be used as coordinates to draw scatter plots in the component space, where the distance between two scores represents their similarity. Two *global factor scores* close to each other indicate high similarity in their respective connectivity patterns across the whole sample, while a *partial factor score* close to its corresponding global factor score represents high similarity between an individual's regional connectivity profile and the regional profile from the whole sample. In sum, covSTATIS provides an unsupervised, linear framework to characterize the similarity among sets of correlation/covariance matrices, and it allows for a one-to-one mapping between input (i.e., whole set of tables and single tables) and output (i.e., *global* and *partial factor scores*). This approach can both identify group-level patterns as well as provide individual-specific expressions of the patterns.

The potential of covSTATIS as a tool for network neuroscience remains largely untapped. Examples of potential applications of covSTATIS include investigations of individual and group differences in spatial and/or temporal network structure in health and disease (**Figure 2, bottom panel**), deep phenotyping of connectivity metrics, and multimodal assessments of network measures within and across individuals (**Figure 2, top panel**). With covSTATIS, multiple types of data tables can now be easily integrated, without requiring *a priori* data simplification, complex black-box implementations, user-dependent specifications or supervised theoretical frameworks. Another promising avenue for covSTATIS is the exploration of brain-behavior relationships within a single framework. Through covSTATIS, participants' correlations—computed from high dimensional brain and behavioral data tables—can be integrated in a unified compromise space from which the shared variance between tables can be extracted. While not exhaustive, the proposed applications of covSTATIS highlight the versatility of the method. We hope that the network neuroscience community will benefit from covSTATIS and collectively further refine and expand the approach. Ongoing developments of covSTATIS can be found on our website: https://giuliabaracc.github.io/covSTATIS_netneuro/.

To facilitate adoption of covSTATIS, we provide a step-by-step tutorial using Open Data[80]. In this tutorial, we use covSTATIS to examine how fMRI-derived functional connectivity reconfigures across task conditions with increasing cognitive load (0-back, 1-back, 2-back), in a healthy adult lifespan sample of 144 individuals. The tutorial can be accessed here: https://giuliabaracc.github.io/covSTATIS_netneuro/pages/tutorial.html. In the tutorial, we additionally guide the reader in the choice of covSTATIS' parameters based on the type of input data.

CovSTATIS serves as a theoretically and computationally accessible tool for similarity analyses, capable of preserving and integrating high dimensional, complex multi-table data typical of network neuroscience. Its linear, unsupervised, user-independent implementation makes covSTATIS a highly interpretable and versatile tool designed to pave the path for new discoveries in network neuroscience.

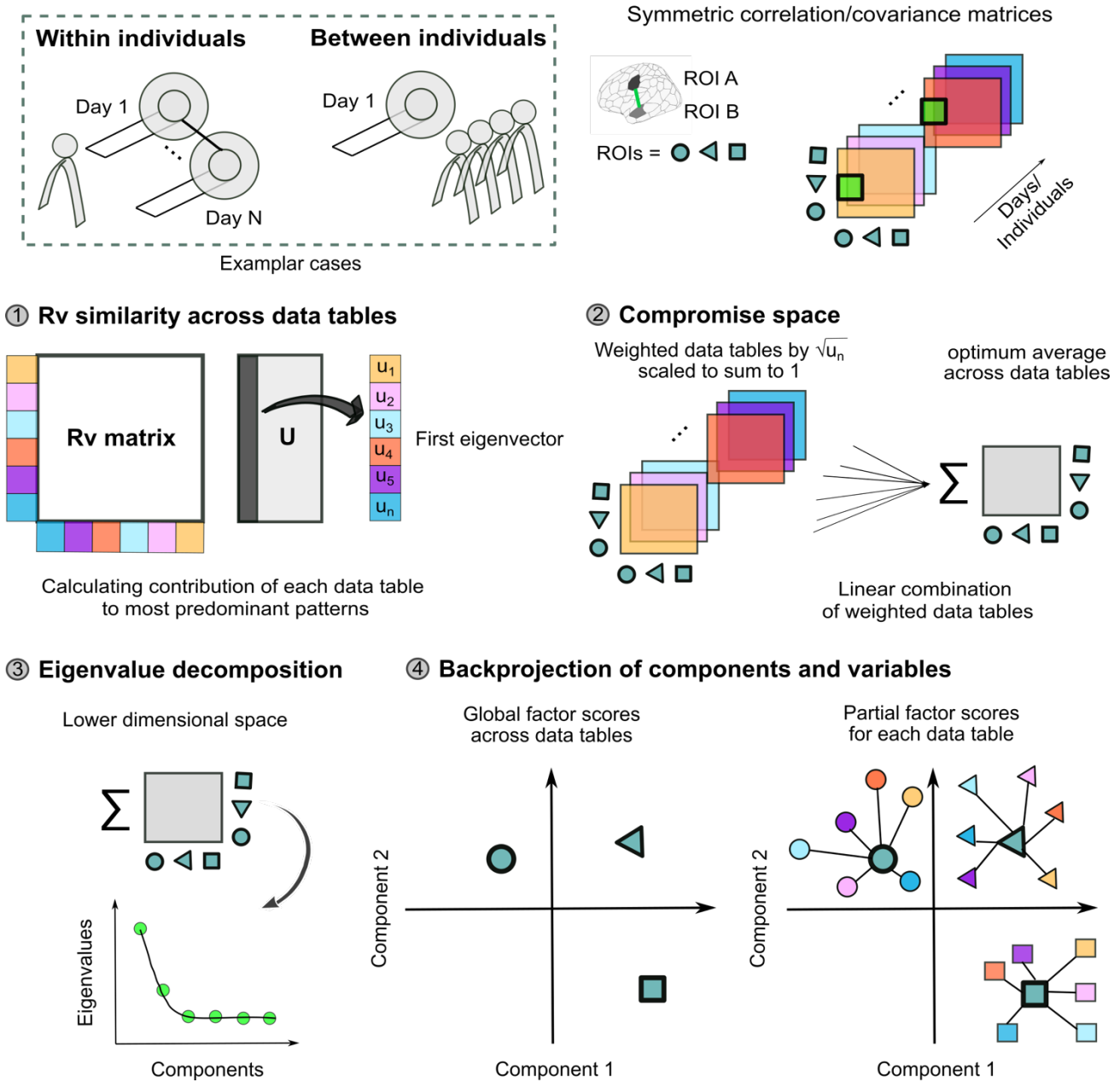

**Figure 1.** CovSTATIS is used to analyze multiple correlation/covariance matrices obtained either within or between individuals. We provide an example using functional connectivity matrices, collected on multiple individuals, as input to covSTATIS. **(1)** First, covSTATIS combines all connectivity matrices by quantifying their overall similarity via their $R_V$ coefficients. These coefficients are then stored in the $\mathbf{R_V}$ matrix. Next, covSTATIS uses the first eigenvector ($\mathbf{u}_1$) of the $\mathbf{R_V}$ matrix to derive weights for each connectivity matrix. **(2)** With these weights, covSTATIS computes the linear combination of all matrices to generate a common space, the *compromise*, which best represents the connectivity pattern across the sample. **(3)** The *compromise* then undergoes eigenvalue decomposition and orthogonal components are extracted to characterize the variance in the whole-sample connectivity pattern. **(4)** The variables of the compromise (illustrated by different shapes of green points; i.e., individual brain regions) are represented as *global factor scores* in the component space. *Global factor scores* represent the connectivity pattern of each brain region across the entire sample. The same variables from each individual matrix can also be back projected onto the same space as *partial factor scores* (indicated by points with the same shape of different colors). *Partial factor scores* represent the connectivity pattern of each brain region for a specific individual. Importantly, the weighted means of all partial factor scores of a given variable equal to their global factor scores (i.e., barycentric property). In this component space, the distance between factor scores provides meaningful and interpretable information about the similarity in the connectivity profile of any two brain regions. The closer the (global or partial) factor scores of two brain regions, the more similar their connectivity profiles.

# covSTATIS: examples of applications in network neuroscience

| Type of research question | Global factor scores | Partial factor scores |
|---|---|---|
| Group comparison/ individual differences | Group means of ROIs/networks | ROI/network configuration per group or individual |
| Deep phenotyping | Session means of ROIs/networks | ROI/network configuration per session |
| Task/condition differences | Tasks/condition means of ROIs/networks | ROI/network configuration per task/condition |
| Multimodal neuroimaging | Modality means of ROI/networks | ROI/network configuration per neuroimaging modality |
| Temporal profiling of network structure (e.g., dynamic functional connectivity) | Means of time chunks (e.g., across sliding windows) of ROI/networks | ROI/network configuration per time chunk (e.g. per window) |

① **Examples of extractable features**

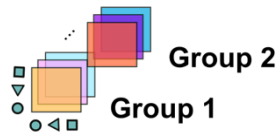

**Group comparison**
If multiple groups, each group can be backprojected separately and partial factor scores can be used to quantify group differences

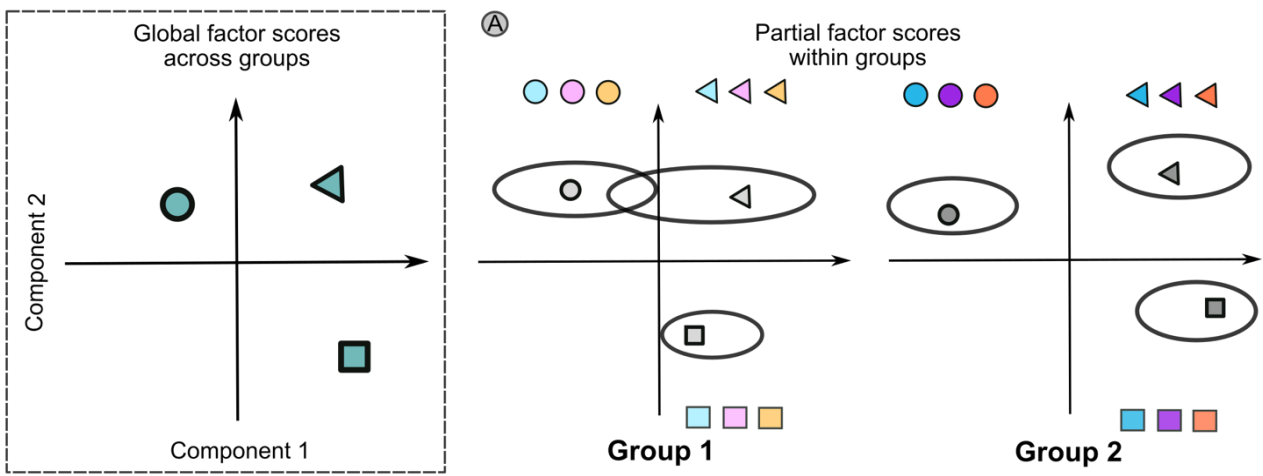

Similarity between observations (across the whole sample or within group) can also be quantified by deriving the area of convex hull

Ⓑ Partial factor scores across groups

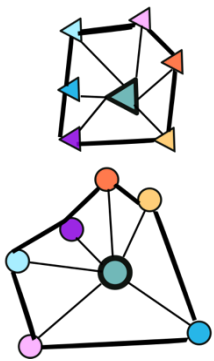

**Smaller area of the hull**
more similar ROI connectivity across (left) or within (right) groups

**Bigger area of the hull**
more diverse ROI connectivity across (left) or within (right) groups

Ⓒ Partial factor scores within groups

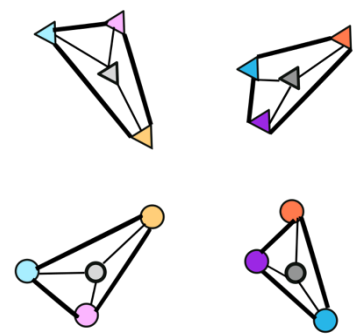

**Figure 2. Top panel:** examples of applications of covSTATIS in network neuroscience. **Bottom panel:** examples of extractable features from covSTATIS. For instance, **(A)** illustrates how we can extract, from global factor scores, group means of partial factor scores, derive their bootstrap confidence intervals, and use them to interpret group differences in network configurations. **(B)** demonstrates how we can quantify the overall heterogeneity among all partial factor scores via computing the area of the hull. **(C)** shows how such heterogeneity can also be evaluated for different groups separately.

## Methods

### Notations

A matrix is denoted by a bold, uppercase letter (e.g., **X**), a vector is denoted by a bold, lowercase letter (e.g., **x**), and an element of a matrix is denoted by a lowercase italic letter (e.g., $x$). The cardinal of a set is denoted by an uppercase italic letter (e.g., $I$). Given $I$ data tables, we used the subscript $i$ to identify individual data tables (e.g., $\mathbf{X}_i$). The boldface capital letter **I** denotes the identity matrix. The transpose of a matrix is denoted by the superscript $^T$ (e.g., $\mathbf{X}^T$).

The $j$th column of matrix **X** is denoted by $\mathbf{x}_j$, and the value on the $k$th row and the $j$th column is denoted by $x_{i,j}$. For an $I \times J$ matrix, the minimum of $I$ and $J$ is the largest possible rank, denoted by $L$, of **X**. The trace(**X**) operator gives the sum of the diagonal elements of the square matrix **X**.

### covSTATIS

To generate the compromise space that best represents the common pattern across all data tables (e.g., correlation/covariance matrices), covSTATIS first derives weights from a pairwise similarity matrix, called the $\mathbf{R_V}$ matrix, which quantifies the similarity between data tables via the $R_V$ coefficient. Formally, given two $J \times J$ data tables $\mathbf{X}_i$ and $\mathbf{X}_{i'}$ (e.g., two connectivity matrices with $J$ ROIs from the 2 observations $i$ and $i'$, e.g., participants or tasks), the $R_V$ coefficient between these two matrices is computed as:

$$R_V = \frac{\text{trace}\left(\mathbf{X}_i^\top \mathbf{X}_{i'}\right)}{\sqrt{\text{trace}\left(\mathbf{X}_i^\top \mathbf{X}_i\right) \text{trace}\left(\mathbf{X}_{i'}^\top \mathbf{X}_{i'}\right)}} = \frac{\sum_{j}^{J} \sum_{j'}^{J} x_{j,j',i}\, x_{j,j',i'}}{\sqrt{\left(\sum_{j}^{J} \sum_{j'}^{J} x_{j,j',i}^2\right)\left(\sum_{j}^{J} \sum_{j'}^{J} x_{j,j',i'}^2\right)}}. \tag{1}$$

Akin to a squared correlation coefficient, the $R_V$ coefficient takes values in the interval [0 1]. The $R_V$ coefficients between all matrices are then stored in an $I \times I$ $\mathbf{R_V}$ matrix, denoted by **C**, where the cell $c_{i,i'}$ stores the value of the $R_V$ coefficient between $\mathbf{X}_i$ and $\mathbf{X}_{i'}$. As **C** gives the similarity between data tables, the first component of **C** best represents the common pattern across tables, and the first eigenvector of **C** ($\mathbf{u}_1$) quantifies how similar each table is to this common pattern. As a result, to build the compromise space, weights for each data table are derived by the scaled $\mathbf{u}_1$, scaled to sum to 1. Formally, $\mathbf{C}$ undergoes EVD:

$$\mathbf{C} = \mathbf{U}\boldsymbol{\Omega}\mathbf{U}^\top \text{ such that } \mathbf{U}^\top\mathbf{U} = \mathbf{I}, \tag{2}$$

where $\boldsymbol{\Omega}$ is an $R \times R$ diagonal matrix of eigenvalues of $\mathbf{C}$ with $R$ denoting the rank of $\mathbf{C}$, and $\mathbf{U}$ is a $I \times R$ matrix of eigenvectors of $\mathbf{C}$. Next the weights of $\mathbf{X}_i$ (denoted by $\alpha_i$) are obtained as:

$$\alpha_i = \sqrt{\frac{u_{i,1}}{\sum_{i=1}^{I} u_{i,1}}} = \left(u_{i,1}(\mathbf{1}^\top \mathbf{u}_1)^{-1}\right)^{\frac{1}{2}}, \tag{3}$$

where $u_{i1}$ is the $i$th value of $\mathbf{u}_1$, which corresponds to $\mathbf{X}_i$. The compromise ($\mathbf{X}_+$) is then computed as the weighted sum of all data matrices, where

$$\mathbf{X}_+ = \sum_{i=1}^{I} \alpha_i \mathbf{X}_i, \tag{4}$$

and decomposed by EVD:

$$\mathbf{X}_+ = \mathbf{P}\boldsymbol{\Lambda}\mathbf{P}^\top \text{ such that } \mathbf{P}^\top\mathbf{P} = \mathbf{I}, \tag{5}$$

where $\boldsymbol{\Lambda}$ is an $L \times L$ diagonal matrix of eigenvalues of $\mathbf{X}_+$ with $L$ denoting the rank of $\mathbf{X}_+$, and $\mathbf{P}$ is a $J \times L$ matrix of eigenvectors of $\mathbf{X}_+$. From EVD, the *global factor scores* $\mathbf{F}$ (i.e., factor scores from the compromise) are computed as:

$$\mathbf{F} = \mathbf{X}_+\mathbf{P}\boldsymbol{\Lambda}^{-\frac{1}{2}} \tag{6}$$

and the *partial factor scores* $\mathbf{F}_i$ (i.e., the factor scores derived from the projection of individual tables onto the compromise) are computed as:

$$\mathbf{F}_i = \mathbf{X}_i\mathbf{P}\boldsymbol{\Lambda}^{-\frac{1}{2}}. \tag{7}$$

It is worth noting that these partial factor scores have a barycentric property, that is their weighted sums equate to the global factor scores:

$$\mathbf{F} = \sum_{i}^{I} \alpha_i \mathbf{F}_i. \tag{8}$$

**The optimization problem in covSTATIS**

Optimization in covSTATIS is a two-part problem. The first part is akin to the optimization problem of principal component analysis, the second part is the same as the optimization problem of eigenvalue decomposition.

*First part*

First, weights for each data table are obtained to compute the compromise, such that the similarity between the compromise and all input matrices is maximal. Second, components are computed that best explain the compromise's inertia (i.e., variance in more than 2 dimensions). Formally, the first optimization problem can be written as the following maximization problem:

$$\arg\max_{\boldsymbol{\alpha}} \sum_{i=1}^{I} \left\langle \mathbf{X}_i, \underbrace{\sum_{i'} \alpha_{i'} \mathbf{X}_{i'}}_{\mathbf{X}_+} \right\rangle^2 \quad \text{with} \quad \boldsymbol{\alpha}^\top \boldsymbol{\alpha} = 1, \tag{9}$$

and the sum of squared scalar products can be developed and simplified:

$$\begin{aligned} \sum_{i=1}^{I} \langle \mathbf{X}_i, \mathbf{X}_+ \rangle^2 &= \sum_{i=1}^{I} \sum_{i'=1}^{I} \alpha_i \alpha_{i'} \sum_{i''=1}^{I} \text{trace}\{\mathbf{X}_i \mathbf{X}_{i''}\} \times \text{trace}\{\mathbf{X}_{i''} \mathbf{X}_{i'}\} \\ &= \sum_{i=1}^{I} \sum_{i'=1}^{I} \alpha_i \alpha_{i'} \sum_{i''=1}^{I} c_{i,i''} \times c_{i'',i'} \\ &= \boldsymbol{\alpha}^\top \mathbf{C}^2 \boldsymbol{\alpha}. \end{aligned} \tag{10}$$

Therefore, the solution of the optimization problem defined by Equation 9 is the first eigenvector of $\mathbf{C}^2$, which is the same as the first eigenvector of $\mathbf{C}$. According to the Perron-Frobenius theorem, the elements of $\boldsymbol{\alpha}$ will all have the same sign (chosen as positive). These elements are then scaled to sum to 1 to ensure that partial factor scores will be barycentric for their respective global factor scores (cf. Equation 8). Note that we ignored the denominator of the $Rv$ in the first line as it is a fixed scalar equal to $J^2$ and has no effect on the maximization problem.

This optimization problem is similar to the optimization problem of Principal Component Analysis (PCA). In PCA, weights are searched *for each variable* to compute factor scores—which are computed as linear combinations of these variables. In covSTATIS, weights are searched *for each data table* to compute the compromise—the linear combination of these data tables.

*Second part*

The second optimization problem is equivalent to the optimization problem of an eigen-decomposition. In the eigen-decomposition, for each component, weights (i.e., loadings) of each row/column are searched to compute factor scores ($\mathbf{F}$). Factor scores are linear combinations of the loadings that have the largest possible variance (as evaluated by their associated eigenvalues). This optimization problem can be written as the minimization problem of approximating the sum of squared factor scores to the compromise:

$$\arg\min_{\mathbf{F}} \|\mathbf{X}_+ - \mathbf{F}\mathbf{F}^\top\|_2^2 \quad \text{such that} \quad \mathbf{F}^\top\mathbf{F} = \mathbf{\Lambda}$$
$$\Leftrightarrow \arg\min_{\mathbf{P}} \|\mathbf{X}_+ - \mathbf{P}\mathbf{\Lambda}\mathbf{P}^\top\|_2^2 \quad \text{such that} \quad \mathbf{P}^\top\mathbf{P} = \mathbf{I}. \tag{11}$$

Here, $\mathbf{P}$ is the matrix of eigenvectors and $\mathbf{\Lambda}$ is the diagonal matrix of eigenvalues.

The detailed proofs and descriptions of the optimization problems of covSTATIS can be found in the Appendix of [64].

## Data availability

Data used in the tutorial are available online (https://osf.io/hnj7s/) and are described in detail in our previous publication[80].

## Code availability

The original source code for covSTATIS can be found here: https://cran.r-project.org/web/packages/DistatisR/ and its helper file here: https://cran.r-project.org/web/packages/DistatisR/DistatisR.pdf. All code used to apply covSTATIS in network neuroscience and replicate our tutorial, along with all documentation, can be accessed here: https://giuliabaracc.github.io/covSTATIS_netneuro/pages/tutorial.html. For the tutorial, a downloadable Rmd file can be accessed here: https://github.com/giuliabaracc/covSTATIS_netneuro/blob/main/pages/tutorial.qmd. The following GitHub (https://github.com/giuliabaracc/covSTATIS_netneuro/tree/main) and website (https://giuliabaracc.github.io/covSTATIS_netneuro/) links serve as centralized resources for covSTATIS' applications in network neuroscience.

## Funding

This research was supported in part by grants from the Canadian Institutes of Health Research (CIHR) and the Natural Sciences and Engineering Research Council of Canada (NSERC). G.B. and R.N.S. are supported in part by Fonds de recherche du Québec (FRQS). J.-C.Y. receives grant support from the Discovery Fund of the Centre for Addiction and Mental Health (CAMH).

## CRediT author statement

**Giulia Baracchini:** Conceptualization, Software, Formal Analysis, Data Curation, Writing – Original Draft, Writing – Review and Editing, Visualization. **Ju-Chi Yu:** Conceptualization, Software, Formal